\documentclass[epj]{webofc-mod}
\usepackage[varg]{txfonts}   % Web of Conferences font
\usepackage[hyperindex,breaklinks]{hyperref}
\usepackage{url} 
\DeclareGraphicsExtensions{.pdf}
\wocname{EPJ Web of Conferences}
\woctitle{ICNFP 2015}
\def\met{\ensuremath{E_{\mathrm{T}}^{\mathrm{miss}}}\,} 

\def\pt{\ensuremath{p_{\mathrm{T}}}\,}

\def\HT{\ensuremath{H_{\mathrm{T}}}\,}

\def\GeV{\ensuremath{\mathrm{GeV}}\,}
\def\TeV{\ensuremath{\mathrm{TeV}}\,}

\begin{document}
\selectlanguage{english}
\hfill IFIC/15-93
\title{Conciliating SUSY with the Z-peaked excess}
%
%   --- 10-12 pages
%

\author{Vasiliki A.\ Mitsou\inst{1}\fnsep\thanks{\email{vasiliki.mitsou@ific.uv.es}}}

\institute{Instituto de F\'isica Corpuscular (IFIC), CSIC -- Universitat de Val\`encia, \\ 
Parc Cient\'ific de la U.V., C/ Catedr\'atico Jos\'e Beltr\'an 2, \\
E-46980 Paterna (Valencia), Spain}

\abstract{%
The ATLAS experiment observed an excess at the $3\sigma$ level in the channel of $Z$~boson, jets and high missing transverse momentum in the full 2012 dataset at 8~TeV while searching for SUSY. The question arises whether the abundance and the kinematical features of this excess are compatible with the yet unconstrained supersymmetric realm, respecting at the same time the measured Higgs boson properties and dark matter density. By trying to explain this signal with SUSY we find that only relatively light gluinos together with a heavy neutralino NLSP decaying predominantly to a $Z$~boson plus a light gravitino could reproduce the excess. We construct an explicit general gauge mediation model able to match the observed signal. More sophisticated models could also reproduce the signal, as long as it features light gluinos, or heavy particles with a strong production cross section, producing at least one $Z$~boson in its decay chain. The implications of our findings for the Run~II at LHC with the scaling on the $Z$ peak, as well as for the direct search of gluinos and other SUSY particles, are also discussed.  
}
\maketitle
\section{Introduction}

Supersymmetry (SUSY)~\cite{susy-rev} is an extension of the Standard Model (SM) that assigns to each SM field a superpartner with a spin differing by half a unit. SUSY  solves in an elegant manner several open issues of the SM, such as the hierarchy problem, the nature of dark matter~\cite{dm-review}, and the grand unification. Although it is one of the most relevant scenarios of new physics explored at the LHC, no clear signs of SUSY have been observed so far. 

Within this scenario, the ATLAS experiment~\cite{Aad:2008zzm} at the LHC~\cite{Evans:2008zzb} observed an intriguing excess of $3\sigma$ in events with $e^+ e^-$ or $\mu^+ \mu^-$ pairs  at the $Z$ peak~\cite{Aad:2015wqa}, accompanied by hadronic activity and missing transverse energy (MET). More precisely, the main selection criteria of the analysis are: (i) At least two same-flavour leptons with opposite electric charge with the leading  (sub-leading) lepton $\pt > 25~\GeV$ ($\pt > 10~\GeV$). Their invariant mass must be within $81 < m_{\ell\ell} < 101~\GeV$; (ii) All events contain at least two jets with $\pt > 35~\GeV$ and $|\eta| < 2.5$, and have $\met > 225~\GeV$ and $\HT \equiv \sum_i \pt^{{\rm jet},i} + \pt^{{\rm lep1}} + \pt^{{\rm lep2}} > 600~\GeV$; (iii) The azimuthal angle between each of the two leading jets and \met\ higher than 0.4.

With an integrated luminosity of $20.3 ~{\rm fb}^{-1}$ of $pp$ collisions at $\sqrt{s}=8~\TeV$, the experiment observes a total of 29~pairs of electrons and muons with an invariant mass compatible with the $Z$-boson mass, with an expected background of $10.6 \pm 3.2$ events\footnote{A similar analysis on $Z$ plus \met has been performed by CMS. However, among other differences, no cut on \HT\ has been applied. No deviation from SM expectations has been observed in that analysis.}. No excess over the expected background is observed outside the $Z$ peak. The question that immediately arises is whether SUSY, or some other extension of the Standard Model (SM), can explain that excess of $Z$+MET events taking into account the current limits on beyond-SM (BSM) physics. A study in those terms within a SUSY framework is presented in Ref.~\cite{Barenboim:2015afa} and the main conclusions are highlighted here.

As we shall see, the observed signal can only be explained if one has a large production cross section of heavy SUSY particles (gluinos or squarks) whose decay chain contains about one $Z$~boson per parent particle. If such an explanation is indeed the answer to the observed excess, our study points out the way to confirm it in the Run~II of LHC, as well as cosmological implications, in particular the particle content of dark matter in the Universe. The resulting scheme of SUSY particle mass hierarchy, including charginos and neutralinos, will be apparent.

%%%%%%%%%%%%%%%%%%%%%%%%%%%%%%%%%%%%%%%%%%%%%%%%%%%%%%%%%%%%%%%%%%%%%%%%%%%%%%%%%%%%%%%%%%%%%%%%%%%
\section{$Z$-boson production in the MSSM}

If the observed excess is confirmed, it would clearly indicate a new non-standard process producing additional $Z$~bosons at LHC energies. $Z$~bosons are regularly produced in the decay chains of most of the SM extensions. Still, this signal would require a significant production of $Z$~bosons without conflicting with all other experimental searches of BSM particles. In fact, using the central value for the expected background and taking into account the BR$(Z\to ee,\mu\mu)$, this would imply that we have produced $273\pm48$ additional $Z$~bosons with $20.3~{\rm fb}^{-1}$. Under the assumption that the $Z$~bosons are produced in the decay chains of generic BSM particles, $Y$, produced in the collision, we need to produce at least $273/{\cal N}(Y\to Z)$ $Y$ particles, with ${\cal N} (Y\to Z)$ the average number of $Z$~bosons produced in the decay of a $Y$ particle. On the other hand, the experimental cuts used in the experiment, namely $n_{\rm jets} \geq 2$, $\met > 225~\GeV$  and $\HT > 600~\GeV$, define further the characteristics of the $Y$ particle and its decays.  

We now follow a bottom-up approach attempting to identify the model within the Minimal Supersymmetric Standard Model (MSSM) framework that may yield the observed $Z$-boson excess\footnote{We assume that the acceptance of the applied selection, also taking into account the reconstruction efficiencies is ideally equal to unity. A ``realistic'' generation of SUSY events using detector simulation taking into account the signal acceptance and detector efficiencies follows later.}. We need 273 (225 at 1$\sigma$) $Z$~bosons if we want to accommodate the observed excess. Assuming that $R$-parity is conserved, supersymmetric particles are produced in pairs in processes of the type $p p \to Y \bar Y$. Thus, the required cross section for this process would be
  \begin{equation}
\sigma( p p \to Y  \bar Y) = \frac{N_{\rm ev}/{\cal N}(Y\to Z)}{\cal L} = \frac{137 (113)/{\cal N}(Y \to Z)}{20.3~ {\rm fb}^{-1} } = \frac{6.7 (5.6)~ {\rm fb}}{{\cal N}(Y\to Z)} \,,
\label{requiredXS}
  \end{equation}
where we take into account that two $Y$ particles are produced in each event.
So, if we obtained one $Z$~boson for each $Y$-particle produced, we would need a production cross section of $6.7 \pm 1.1~{\rm fb}$ at the LHC at $\sqrt{s} = 8~\TeV$. Here, we consider the production cross sections of different supersymmetric particles separately to identify the relevant $Z$-producing processes. However in the next section, by employing full event generation, the production of different sparticles contributes to the final $Z$ plus jets plus MET signal.  

Naively, the first option to consider in a hadron collider would be strong production of squarks or gluinos (assuming they produce $Z$ bosons in their decays). However, current experimental searches of jets plus missing energy at LHC force the masses of these coloured particles to be high \cite{Khachatryan:2015pwa,Aad:2015mia,Aad:2014wea}. Nevertheless, as we will see below, in some cases we can still find cross sections of the required size. 

\begin{figure}[ht]
  \centering
  \includegraphics[width=0.47\linewidth]{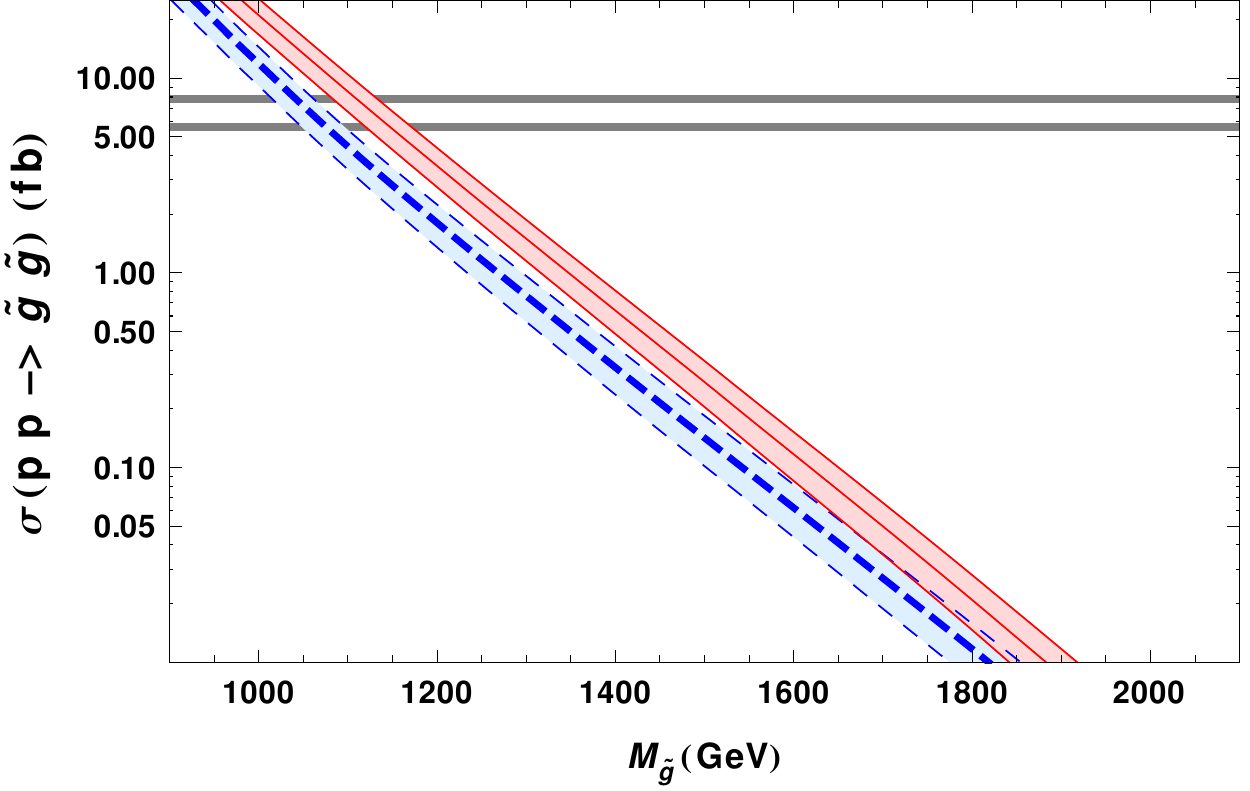} 
  \hfill
  \includegraphics[width=0.47\linewidth]{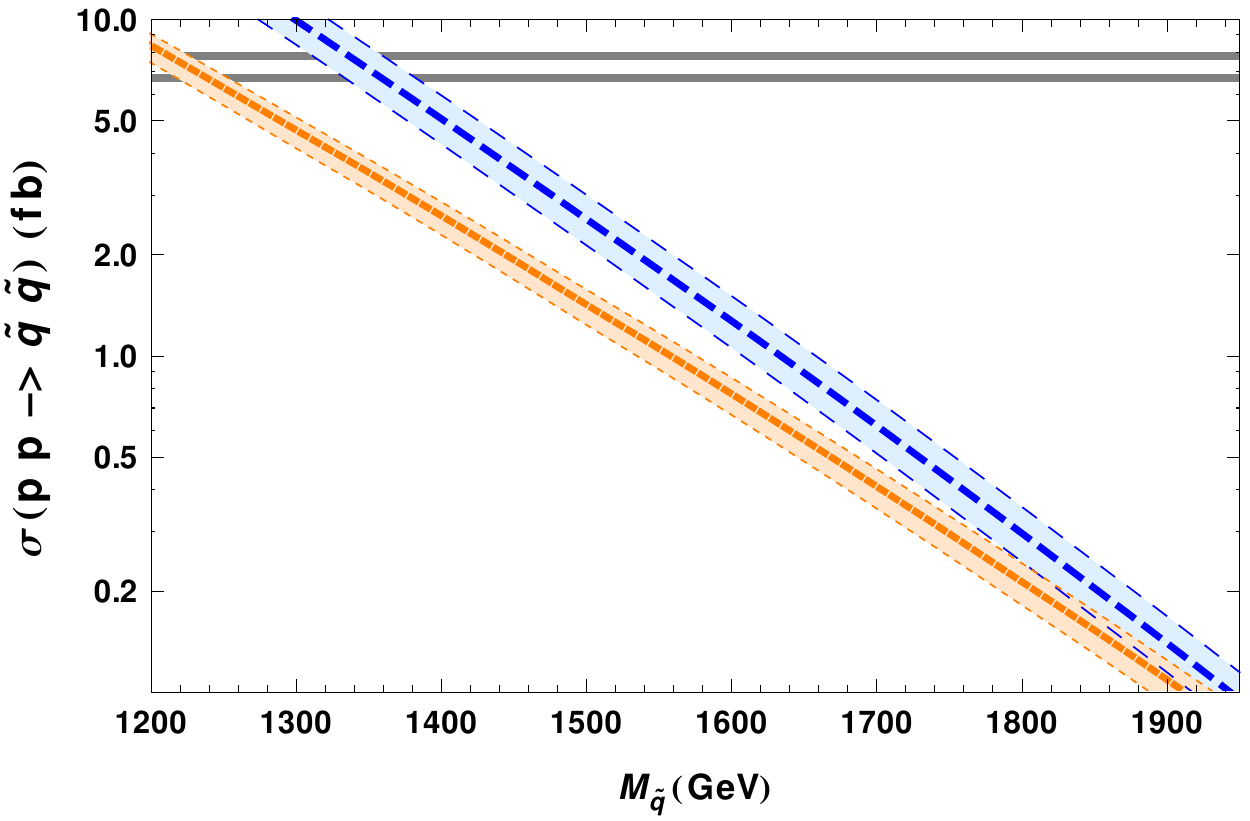}
  \caption{\emph{Left:} Production cross section of gluino pairs as a function of the gluino mass for two fixed values of the first generation squark masses: 1000~\GeV\ (dashed blue) and decoupled squarks (solid red). \emph{Right:} Production cross section of squarks pairs as a function of their mass with gluino masses of 1000~\GeV\ (dashed blue) and 2500~\GeV\ (dotted orange). The bands correspond to $\pm1\sigma$ theoretical uncertainty. The horizontal gray bands indicate the experimental requirements as calculated in Eq.~(\ref{requiredXS}). From Ref.~\cite{Barenboim:2015afa}.}
\label{xsgluqq}
  \end{figure}

Production cross sections of gluinos and squarks depend only on their masses and are basically independent of other MSSM parameters. In the case of gluino and squarks of the first generation, the cross section depends both on the squark and gluino masses due to $t$-channel contributions, but in the case of stops or sbottoms it depends only on the stop or sbottom  mass. In Fig.~\ref{xsgluqq} we present the production cross section of gluino pairs (left) and light-flavour squark pairs (right) calculated at NLL+NLO as a function of the gluino or squark mass. 

As we can see in these figures, the required cross section is reached only for light gluino and squark masses. In the case of gluino production, the needed cross section is obtained only for $m_{\tilde g} \lesssim 1200~\GeV$  and favours heavy squark masses. In fact, these gluino masses are in the boundary of the allowed region obtained from jets plus missing-$E_T$ searches at LHC~\cite{Khachatryan:2015pwa,Aad:2015mia,Aad:2014wea} and would contribute significantly only if every $\tilde g$ produces at least a $Z$~boson in its decay. For the production of squark pairs, present limits are $m_{\tilde q} \gtrsim 1400~\GeV$ for heavy gluinos and  $m_{\tilde q} \gtrsim 1650~\GeV$ for degenerate squarks and gluinos. Under these conditions, $\sigma ( p p \to \tilde q \tilde q)$ is always well below the required cross section, even for $m_{\tilde q} \gtrsim 1400~\GeV$. 

Another important process is the simultaneous production of squark and gluino. However,  we found that we would need both squark and gluino to be light, which is not possible if we take into account the bounds from LHC searches~\cite{Khachatryan:2015pwa,Aad:2015mia,Aad:2014wea}. In summary, the best option seems to be gluino pair production with $m_{\tilde g} \lesssim 1200~\GeV$ with relatively heavy squarks  $m_{\tilde q} \gtrsim 3000~\GeV$, if we can get at least one $Z$ boson in every gluino decay.  

\begin{figure}[ht]
  \centering
  \includegraphics[width=0.47\linewidth]{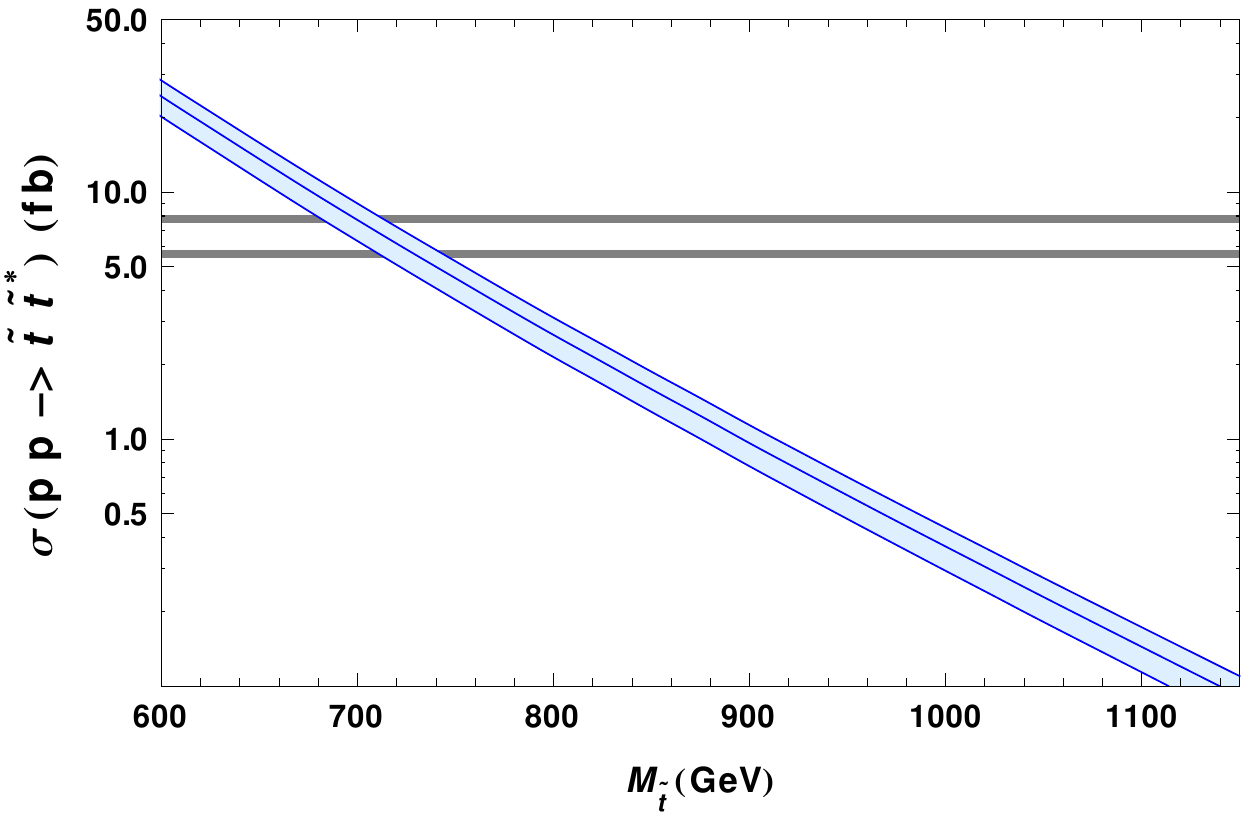}
  \hfill
  \includegraphics[width=0.47\linewidth]{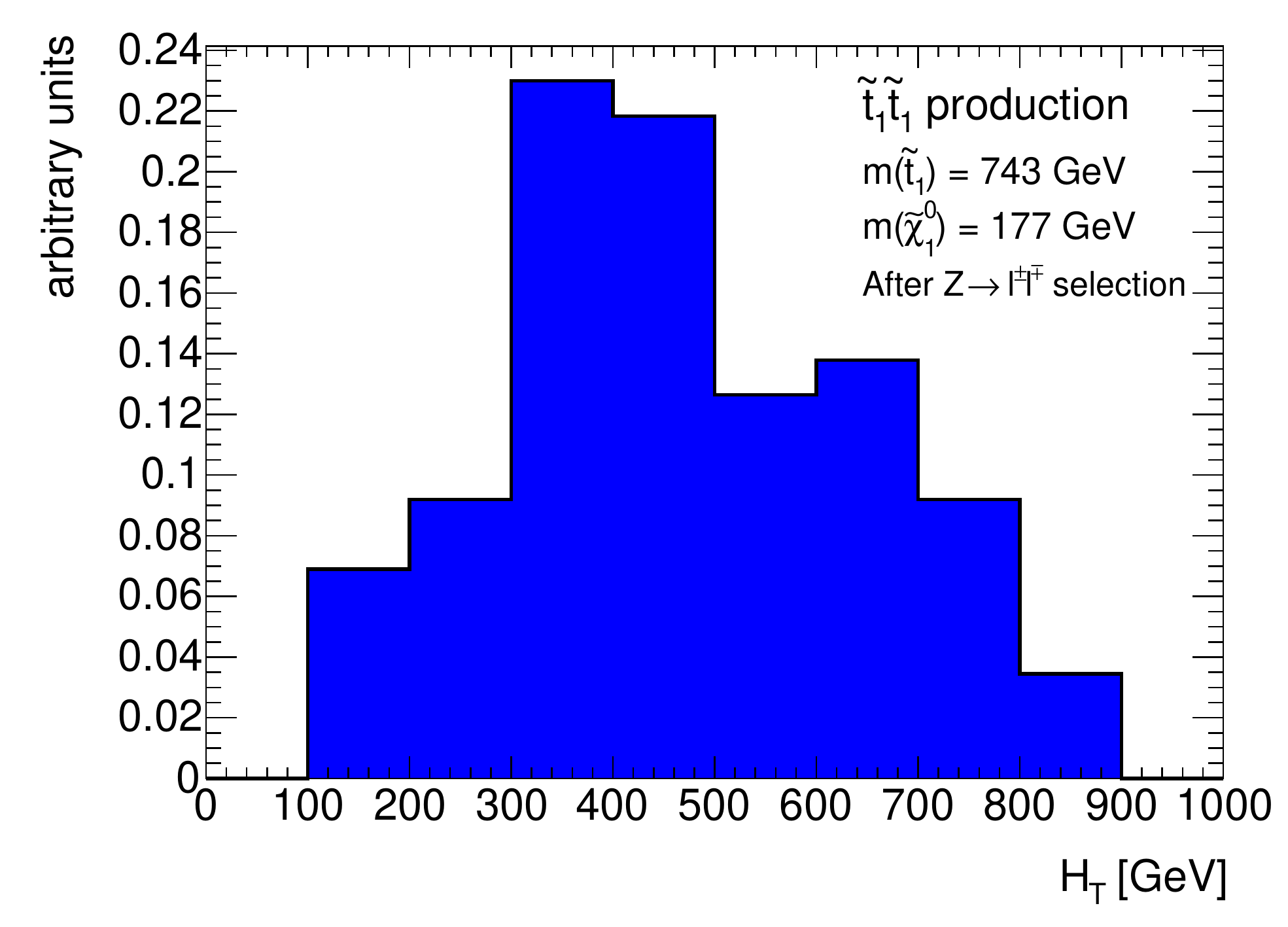}
  \caption{\emph{Left:} Production cross section of $\tilde t \tilde t^*$ pairs as a function of the stop mass. The cross section is practically independent of other sparticle masses. \emph{Right:} \HT\ distribution in arbitrary units from the decay of a pair of stops of $m_{\tilde t} =743$~GeV to jets plus \met after applying the selection $Z\to \ell^+ \ell^-$. From Ref.~\cite{Barenboim:2015afa}.}
\label{xstt}
  \end{figure}

We can also consider the strong production of stop pairs, where the current bounds on stop masses are much lower, $m_{\tilde t} \gtrsim 650~\GeV$ \cite{Khachatryan:2014doa,Aad:2014kra}. The total  $\tilde t \tilde t^*$ production cross section is shown in the left panel of Fig.~\ref{xstt}, calculated at NLL+NLO. In this case, we can see that we could reach the required cross section for $m_{\tilde t} \lesssim 750~\GeV$ which in principle could be achievable in general SUSY models (always assuming that every stop produces a $Z$~boson in its decay). However, the cuts $\HT > 600~\GeV$ and $\met > 225~\GeV$ are very restrictive. We can see this in the right panel of Fig.~\ref{xstt}, where we show the $\HT$ distribution from the decays of stop pairs with  $m_{\tilde t_1}= 750~\GeV$. As can be seen here, the $\HT$ distribution peaks at $\HT \simeq 400-500~\GeV$ as a consequence of the relatively small stop mass, and only a small fraction of the events are able to overcome the cut on $\HT$.  Therefore, we must conclude that it is not possible to generate the required cross section and fulfil the requirements of the observed excess through stop production.  
  
Apart from the production cross sections of coloured sparticles, we also considered weak production of charginos/neutralinos which can be large enough for light gauginos.  Taking into account that the current bounds on chargino and neutralino masses are not very stringent~\cite{Khachatryan:2014mma,Aad:2014vma}, electroweak production is worth exploring. In terms of gauge eigenstates the largest electroweak production cross sections are those corresponding to $\tilde W ^0 \tilde W^\pm$ and $\tilde W ^+ \tilde W^-$, corresponding to $\tilde{\chi}^0_2 \tilde{\chi}_1^\pm$ and  $\tilde{\chi}^+_1 \tilde{\chi}_1^-$ in terms of mass eigenstates. In principle, the production cross section of charginos of $m_{\tilde\chi^+_1} \lesssim 350~\GeV$ appears to be sufficient to accommodate the required $Z$ production if the only restriction imposed is a minimum $\pt \geq 20~\GeV$ for the jets. The required $\met > 225~\GeV$ is easily obtained if $m_{\tilde \chi^0_1} \gtrsim 150~\GeV$, however the high-\HT\ requirement proves to be too restrictive for the weak production. 

Therefore we conclude that, although electroweak production could contribute efficiently to the production of additional $Z$~bosons, these events can not overcome the experimental cuts and can not give rise to the observed signal. With this channel, we have reviewed all relevant production cross sections of different supersymmetric particles that could potentially explain the signal. The next step would be to calculate the average number of $Z$~bosons per parent particle $Y$, that we use in Eq.~(\ref{requiredXS}).

%%%%%%%%%%%%%%%%%%%%%%%%%%%%%%%%%%%%%%%%%%%%%%%%%%%%%%%%%%%%%%%%%%%%%%%%%%%%%%%%%%%%%%%%%%%%%%%%%%%
\section{Decay of sparticles to $Z$~bosons}

$Z$~bosons are produced through the decay chains of most MSSM particles, although the multiplicity obtained per each supersymmetric particle depends on the identity of the supersymmetric particle initially produced and on the supersymmetric spectrum below its mass. The main sources of $Z$~bosons are decays of neutralinos and charginos and also some squarks. 
  
We can obtain $Z$~bosons in the decays of higgsino-like neutralinos and charginos. For instance in a usual minimal Supergravity (mSUGRA) spectrum, the second neutralino will only produce $Z$~bosons through its (relatively small) higgsino component while the two heavier neutralinos can be expected to produce a sizeable number of $Z$~bosons. On the other hand, charginos can produce $Z$~bosons both through the wino and from the higgsino component but only in decays of the heavier charginos, as the lightest one will only decay to a $W$~boson and a neutralino (or lepton-slepton if $m_{\tilde \ell}\leq m_{\tilde\chi^+}$).

\begin{figure}[ht]
  \centering
  \includegraphics[width=0.47\linewidth]{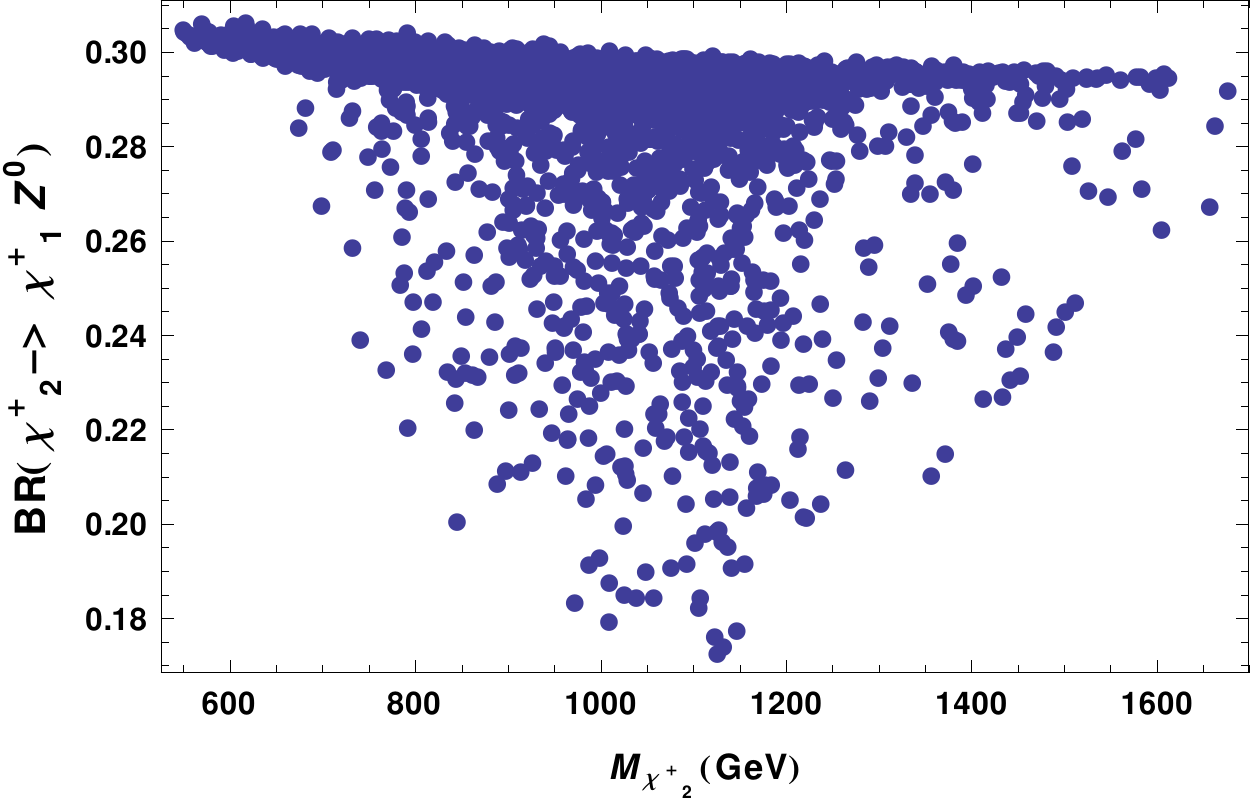}
  \hfill
  \includegraphics[width=0.47\linewidth]{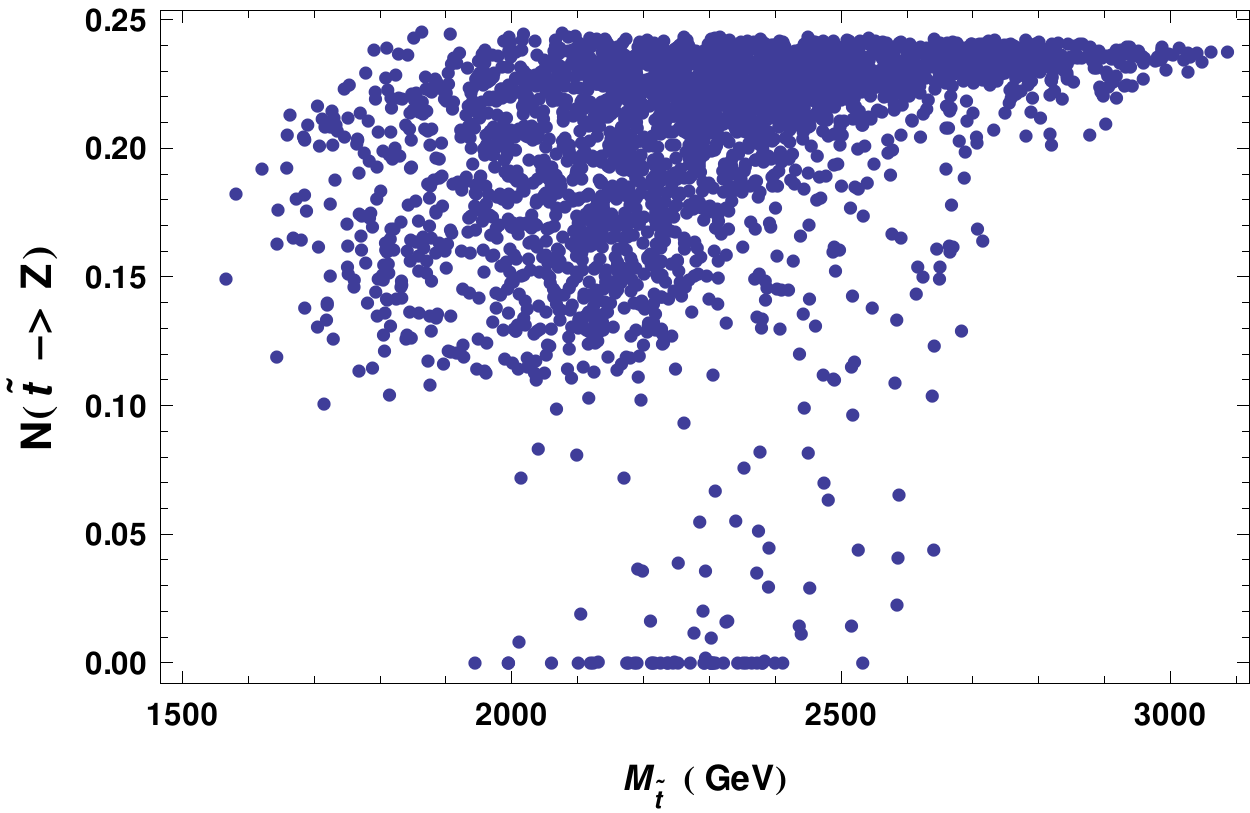}
  \caption{Average number of $Z$ bosons, ${\cal N}(Y \to Z)$ in decays of $\tilde\chi^+_2$ (left) and $\tilde t$ (right) for a typical mSUGRA spectrum. From Ref.~\cite{Barenboim:2015afa}.}
\label{BRchip}
  \end{figure}

Let us examine the case of the heaviest electroweak neutralinos. We obtain ${\cal N}(\tilde\chi_2^0 \to Z)$ around 0.1, as expected if the higgsino content is relatively small. Then, ${\cal N}(\tilde\chi_3^0 \to Z)$ can reach at most 0.45 while the other 50\% of the decays go to $\tilde\chi^\pm W^\mp$. Similarly, the heavy charginos produce $Z$~bosons in their decay as can be seen in the left panel of Fig.~\ref{BRchip}. In this case ${\cal N}(\tilde\chi^+_2 \to Z)$ can be 0.3, while we have similar branching ratios to $\tilde\chi_1^0 W^+$ and $\tilde\chi_1^+ h$. 

Besides chargino and neutralino decays, $Z$~bosons couple also to sfermions through couplings that are chirality diagonal and, therefore in decays, they can only enter through chirality mixing. Although these couplings could be flavour changing, flavour mixing is bound to be small due to the stringent flavour changing neutral current constraints. Therefore we can expect a sizeable amount of $Z$~bosons produced only through chirality mixing in third generation sfermions in decays like $\tilde t_2 \to \tilde t_1 Z$, or $\tilde b_2 \to \tilde b_1 Z$ in the large $\tan \beta$ regime. 

In addition, we obtain $Z$~bosons in the decay chains of strongly-produced sparticles. We can obtain $Z$~bosons at different steps of the decay chain, either through the couplings of $Z$ to sfermions or to charginos/neutralinos that we saw above. For instance, if we produce a pair of gluinos a possible decay chain would be $\tilde g \to \tilde t_2 t \to \tilde t_1 Z~ t \to \tilde\chi_2^+ b~ Z~ t \to
\tilde\chi_1^+ Z~ b~ Z~ t \to \tilde\chi_1^0 W^+~Z~ b~ Z~ t$. Therefore taking into account the corresponding branching ratios, this decay chain would contribute with $2 \times \mbox{BR}(\tilde g \to \tilde\chi_1^0 W^+~Z~ b~ Z~ t)$ to
${\cal N}(\tilde g \to Z )$, the number of $Z$~bosons produced per $\tilde g$ produced.

As we can see in the right panel of Fig.~\ref{BRchip}, in a typical MSSM spectrum we obtain at most 0.2 $Z$~bosons per stop (or gluino) while other squarks produce far fewer $Z$~bosons per squark. Although the plots of Fig.~\ref{BRchip} have been obtained from a mSUGRA spectrum, the expected number of $Z$~bosons would be very similar in other MSSM versions, as it depends only on the spectrum below the mass of the originally produced $Y$ particle and the content of charginos/neutralinos. 

As shown in the previous section, the only possibility to explain the signal is to produce a stop or a gluino-pair as the lightest coloured sparticle, being all other squarks much heavier and only neutralinos, charginos and possibly some sleptons can be below the gluino or stop mass. Moreover, given the size of production cross sections consistent with the present searches, we need to obtain nearly one $Z$~boson per $Y$ particle produced. Therefore, from the expected numbers of $Z$~bosons that we have seen in this section, we have to conclude that it is not possible to reproduce the observed signal in a MSSM with a stable (and light) neutralino. 
 
Nevertheless, we can still consider different variations of the MSSM: 
\begin{itemize}
\item 
A first possibility would be to have a light gluino below 1~\TeV\ that can evade the bounds from jets plus missing $E_T$ if it decays to a sufficiently heavy neutralino LSP in a sort of compressed spectrum. Under these conditions the gluino would be abundantly produced at LHC and even a small number of $Z$~bosons per gluino could fulfil the requirements to explain the observed signal. However, this would require strongly non-universal gaugino masses and very heavy LSPs.
\item A second option is to consider an MSSM where the lightest neutralino decays to a lighter gravitino plus some a $Z$~boson. This is the case in gauge mediated MSSM~\cite{Dine:1993yw,Dine:1994vc,Dine:1995ag,Giudice:1998bp} and it could be also possible in gravity mediated MSSM if the gravitino is lighter than the neutralino which then becomes the NLSP~\cite{Feng:2004mt}. In this case, the neutralino decays to $Z$~boson and gravitino if it is allowed by phase space and the branching ratio will depend on the lightest-neutralino composition. This is the possibility we will explore in the following.
\end{itemize} 
Thus, we analyse a situation where the neutralino is the NLSP and the LSP is the gravitino. All supersymmetric particles will decay to the lightest neutralino which then decays to gravitino plus a photon, a $Z$~boson or a Higgs. The decay width of the lightest neutralino to photon, $h$ or $Z$ plus gravitino~\cite{Moroi:1995fs,Ellis:2003dn,Feng:2004mt} is given by
\begin{eqnarray}
\qquad\quad\Gamma(\tilde\chi_1^0 \to \gamma \tilde G) &=& \frac{\left|N_{11} \cos \theta_W +N_{12} \sin \theta_W\right|^2}{48\pi M_{Pl}^2} \frac{m_{\tilde\chi}^5}{m_{\tilde G}^2} \left[1 -\frac{m_{\tilde G}^2}{m_{\tilde\chi}^2}\right] \left[1 +3\frac{m_{\tilde G}^2}{m_{\tilde\chi}^2}\right] \, , \nonumber \\
\qquad\quad\Gamma(\tilde\chi_1^0 \to Z \tilde G) &=& \frac{\left|-N_{11} \sin \theta_W +N_{12} \cos \theta_W\right|^2}{48\pi M_{Pl}^2} \frac{m_{\tilde\chi}^5}{m_{\tilde G}^2} F(m_{\tilde\chi},m_{\tilde G},m_Z)\, , \nonumber \\
\qquad\quad\Gamma(\tilde\chi_1^0 \to h \tilde G) &=& \frac{\left|-N_{13} \sin \alpha +N_{14} \cos \alpha\right|^2}{96\pi M_{Pl}^2} \frac{m_{\tilde\chi}^5}{m_{\tilde G}^2} F(m_{\tilde\chi},m_{\tilde G},m_h)\, ,
\end{eqnarray}
with $F(x,y,z)$ a function of the particle masses not relevant to our discussion that can be obtained from Ref.~\cite{Feng:2004mt}.
As we can see, if the lightest neutralino is bino-like, $N_{11}\simeq 1$, and the mass difference between neutralino and gravitino is larger that the $Z$~mass, the branching ratios are BR$(\tilde\chi_1^0 \to \tilde G \gamma )\simeq \cos^2 \theta_W \simeq 0.8$ and  BR$(\tilde\chi_1^0 \to \tilde G $Z$ )\simeq \sin^2 \theta_W \simeq 0.2$. Similarly if the lightest neutralino is wino-like, the branching ratios get exchanged. From this equation we can also see that it is possible to get a very large BR to $Z$ bosons as needed to reproduce the observed signal if $(-N_{11} \sin \theta_W +N_{12} \cos \theta_W) \simeq 1$, but this is only possible if the lightest neutralino has a very large wino component.

Although in gauge mediation models the LSP is always the gravitino, the gaugino masses in minimal models are proportional to the gauge couplings and therefore the LSP is mostly bino with small wino and higgsino components. Then, we will have to consider other extensions of the gauge mediation idea, which is the subjecy of the next section.

%%%%%%%%%%%%%%%%%%%%%%%%%%%%%%%%%%%%%%%%%%%%%%%%%%%%%%%%%%%%%%%%%%%%%%%%%%%%%%%%%%%%%%%%%%%%%%%%%%%
\section{A possible explanation in General Gauge Mediation}\label{GGM}

Minimal gauge mediation predicts that all scalar and gaugino masses originate from a single scale and powers of the gauge couplings~\cite{Giudice:1998bp}. Recently a model-independent generalisation of gauge mediation was proposed under the name of General Gauge Mediation (GGM)~\cite{Meade:2008wd,Buican:2008ws}, where all the dependence of soft masses on the hidden sector is encoded in three real and three complex parameters obtained from a small set of current-current correlators. In these models the gaugino and sfermion masses are given by the relations
 \begin{eqnarray}
\qquad\qquad\qquad\qquad\qquad M_r &=& g_r^2 M_s \tilde B_r^{1/2}(0) \\
\qquad\qquad\qquad\qquad\qquad m^2_{\tilde f} &=& g_1^2 Y_f \zeta + \sum_{r=1}^3 g_r^2 {\cal C}_2(f|r) M_s^2 \tilde A_r \nonumber \,,
  \end{eqnarray}
with
\begin{equation}
\tilde A_r = -\frac{1}{16 \pi^2} \int dy \left( 3 \tilde C_1^{(r)}(y) -4 \tilde C_{1/2}^{(r)}(y)+ \tilde C_0^{(r)}(y) \right)\,,
\end{equation}
$\tilde B_r^{1/2}(0)$, $ C_\rho^{(r)}(y)$ (with $\rho=0,1/2,1$, corresponding to scalar, fermion and vector) are associated with the current-current correlators in the hidden sector, $\zeta$ is a possible Fayet-Illiopoulos term ($\zeta=0$ in the following), ${\cal C}_2(f|r)$ the quadratic Casimirs and $M_s$ a characteristic SUSY-breaking scale in the hidden sector.

Having six parameters, $ \tilde B_r^{1/2}(0)$ and $\tilde A_r$, to fix the soft masses in the observable sector, it is clear now that we have much more freedom in GGM \cite{Carpenter:2008he,Rajaraman:2009ga,Thalapillil:2010ek,Kats:2011qh} and, in particular, we have
\begin{equation}
\frac{M_1}{g_1^2} \neq \frac{M_2}{g_2^2} \neq  \frac{M_3}{g_3^2} \,,
\end{equation}
as required to reproduce the observed signal at ATLAS. In particular, we need the NLSP to decay to gravitino and a $Z$~boson with a branching ratio close to one. Fortunately, this is possible in GGM as shown in Ref.~\cite{Ruderman:2011vv,Kats:2011qh}.

In this GGM scenario, the spectrum calculator \textsc{SPheno}~\cite{Porod:2003um,Porod:2011nf} is used to obtain the full supersymmetric spectrum at LHC energies. We define the GGM1 parameter point with the following parameters, $M_s = 400$~TeV, $\tilde B_1^{1/2}=\tilde A_1=309$~TeV, $\tilde B_2^{1/2}=\tilde A_2 =151$~TeV, $\tilde B_3^{1/2}=129$~TeV, $\tilde A_3=316$~TeV and $\tan\beta=9.8$. With these parameters we obtain the spectrum shown in Table \ref{tab1}.

\begin{table}[ht]
\centering
\caption{SUSY spectrum in the GGM1 parameter point. From Ref.~\cite{Barenboim:2015afa}.}\label{tab1} 
\begin{tabular*}{0.95\textwidth}{@{\extracolsep{\fill}}l c c c c c c c c}
\hline
{\rm Particle}& $\tilde g$& $\tilde\chi_1^0$ & $\tilde\chi_2^0$ & $\tilde\chi_3^0$  & $\tilde\chi_4^0$ & $\tilde\chi_1^\pm$ & $\tilde\chi_2^\pm$& $\tilde G$ \\
{\rm Mass [GeV]} & 1088.0 & 428.4 & 431.34 & 1357.0 & 1360.9 & 429.1 & 1361.0 & $4.8 \times 10^{-9}$ \\
\hline 
{\rm Particle}&  $\tilde q_L$ & $\tilde q_R$ & $\tilde b_1$ & $\tilde b_2$ & $\tilde t_1$ &  $\tilde t_2$ & $\tilde \ell_L$& $\tilde \ell_R$ \\
{\rm Mass [GeV]} & 3006 & 2957 & 2876 & 2952 & 2716 & 2881 & 5863 & 5328 \\
\hline
{\rm Particle}&  $h$ & $H$ & $A$ & $H^+$ & & & & \\
{\rm Mass [GeV]} & 119.4 & 1471 & 1471 & 1473 & & & & \\
\hline
\end{tabular*}
\end{table}

With respect to this spectrum, some comments are in order: 
\begin{enumerate} 
\item The two lightest neutralinos and the lightest chargino are very similar in mass, $\sim 430~\GeV$ and this allows a large neutrino mixing as required. In fact the neutralino mixing matrix is given by
\begin{eqnarray}  
\label{Nmix}
\qquad\qquad N_{ij}\simeq\left( 
\begin{array}{cccc}
-0.51 & 0.85& -0.076& 0.031 \\ 
0.86 & 0.51& -0.0024& 0.0071 \\ 
-0.015 & 0.028& 0.71& 0.71\\
-0.037 & 0.065& 0.70& -0.71
\end{array}%
\right)
\end{eqnarray}
On the other hand the relatively large NLSP mass is needed to overcome the \met cut.
\item The gluino is relatively light $m_{\tilde g} = 1088.0~\GeV$ which allows
 for a sizeable production cross section and taking into account the squark masses of order $\sim 3~\TeV$, this mass is allowed by the latest LHC bounds.
\item The lightest Higgs mass must reproduce the observed value at LHC of $m_h\simeq 125~\GeV$ and in this spectrum it reaches only $119.4~\GeV$. This problem (typical in minimal gauge mediation models) can be solved either by increasing the stop masses taking a larger $\tilde A_3$ or assuming extra operators in the Higgs sector, as the dimension-5 operators proposed in Ref.~\cite{Dine:2007xi}. Here, we assume that this problem is solved by one of these mechanisms, given it does not affect the observed phenomenology on the $Z$~peak. 
\end{enumerate}

  \begin{figure}[ht]
  \centering
  \includegraphics[width=0.48\linewidth]{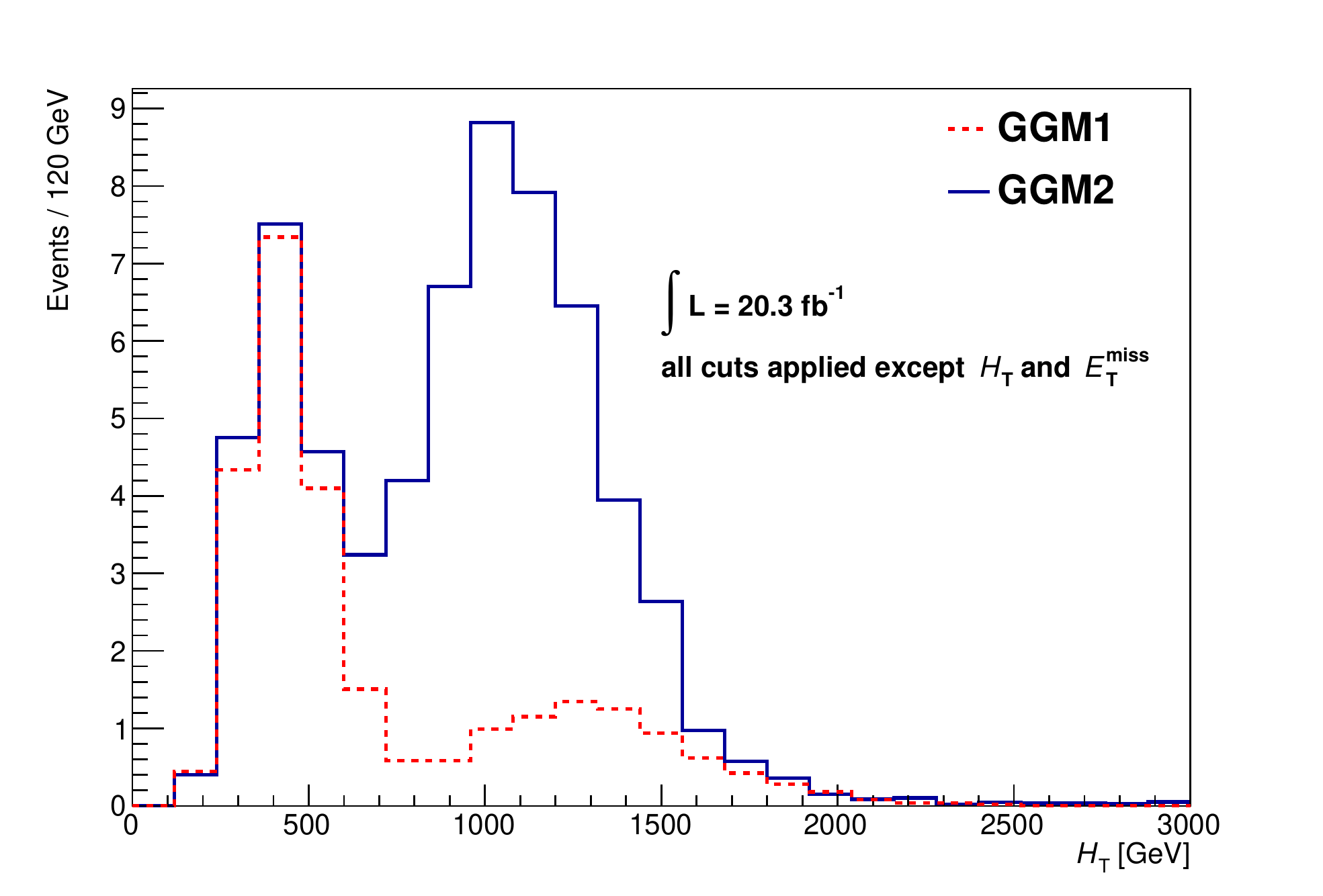} 
  \hfill
  \includegraphics[width=0.48\linewidth]{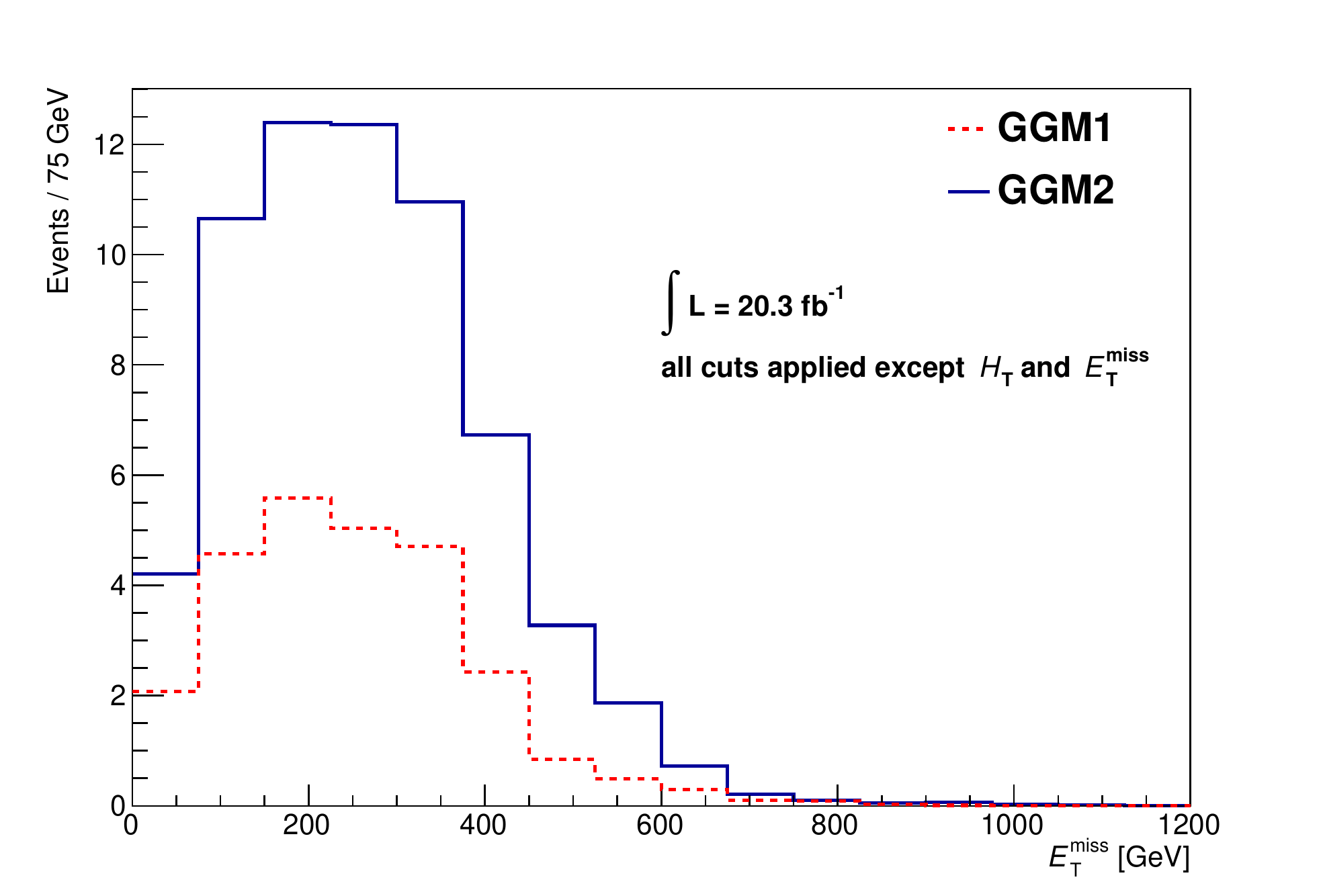}
  \caption{$\HT$ (left) and \met (right) distributions after applying the selection criteria of the Z+MET analysis~\cite{Aad:2015wqa}, except for the cuts on \HT and \met, for the GGM1 (dashed red) and the GGM2 point (solid blue). From Ref.~\cite{Barenboim:2015afa}.}
  \label{GGMHTMET}
  \end{figure}

  \begin{figure}[ht]
  \centering
  \sidecaption
  \includegraphics[width=0.48\linewidth]{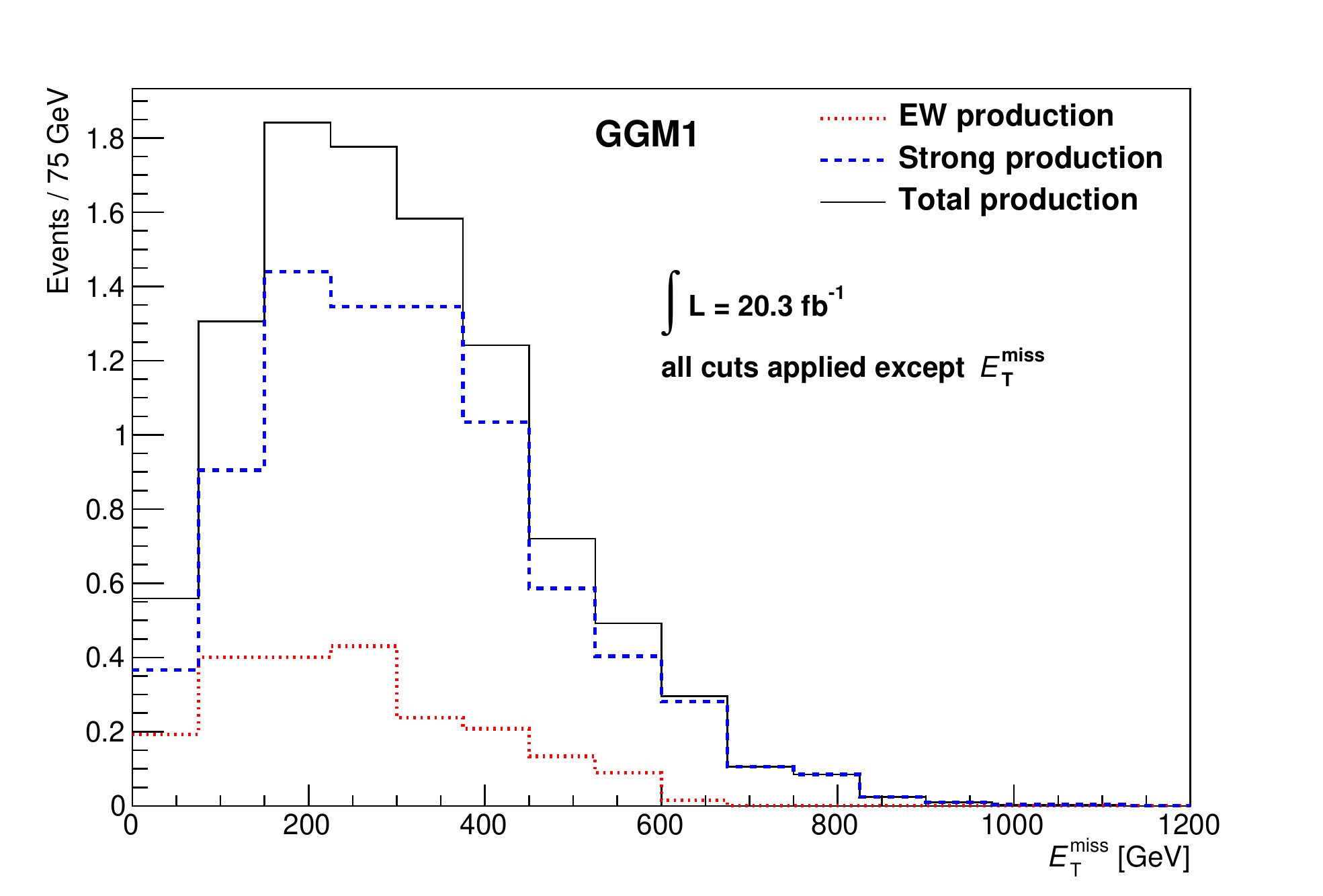} 
  \caption{\met distribution corresponding to the GGM1 point from strong production (solid blue) and electroweak production (dashed red) with the same selection as in Fig.~\ref{GGMHTMET}, after applying the $\HT>600~\GeV$ cut. From Ref.~\cite{Barenboim:2015afa}.}
  \label{GGM1}
  \end{figure}

Under these conditions, the lightest neutralino width is $\Gamma_{\tilde\chi_1^0} = 4.097 \times 10^{-10}~\GeV$ and the decay branching ratios are BR$(\tilde\chi^0_1 \to \tilde G \gamma)=1.14 \times 10^{-3}$,  BR$(\tilde\chi^0_1 \to \tilde G Z)= 0.997$ and  BR$(\tilde\chi^0_1 \to \tilde G h)=1.35 \times 10^{-3}$. Therefore, gluinos are produced at LHC with a cross section of $(8.4 \pm 1.6)$~fb at NLL+NLO and after going through different decay chains all of them produce a $Z$~boson plus a gravitino. In this case, the strong production represents approximately $20\%$ of the total production cross section.

We simulate the production of supersymmetric particles at LHC at 8~\TeV\ (LHC8) with this spectrum using \textsc{Pythia8}~\cite{0710.3820} and \textsc{Prospino2}~\cite{Beenakker:1996ch,Beenakker:1997ut,Beenakker:1999xh} $K$-factors and the response of the ATLAS detector using \textsc{Delphes}~\cite{deFavereau:2013fsa}. 
The selection of events for this study is performed as close as possible to the search performed in ATLAS~\cite{Aad:2015wqa}. The dashed red line in Figs.~\ref{GGMHTMET} and~\ref{GGM1} shows the \HT\ and \met\ distributions respectively for the GGM1 point after applying all the aforementioned selection cuts except for the \HT\ and \met\ cuts. 
In the \HT\ distribution, we can distinguish the two peaks corresponding to electroweak production at low \HT\ values and gluino production at higher \HT. From here, we can expect that the cut on \HT\ will eliminate most of the electroweak production but not the gluino production.
This can be seen in Fig.~\ref{GGM1}, where the \met\ distribution is presented separately for strongly produced events (solid blue line) and for the electroweak component of the production (dashed red line) for the same selection as in Fig.~\ref{GGMHTMET} but after applying the \HT\ cut, i.e. the final selection except for the cut on \met. The electroweak component is significantly reduced by the \HT\ cut while mainly only events coming from strong production survive the cut, as expected.  In fact, in this simulation of point GGM1, 99\% of the gluino points and only 11\% of the electroweak points have survived the \HT\ cut.    

We see that the peak in the \met\ distribution is approximately at $m_{\tilde\chi_1^0}/2$, and, for $m_{\tilde\chi_1^0}= 425~\GeV$, a reasonable fraction of the events will survive the \met\ cut at 225~\GeV. In the simulation, 65\% of the gluino point and 53\% of the electroweak points survive this cut. However, due the relatively small production cross section, the final number of events is small. 
In this simulation and after applying all relevant experimental cuts, an expected signal of $6.34 \pm 1.02$ lepton pairs is found, to be compared with the observed excess of  $19.4 \pm 3.2$. This number of surviving events was obtained at NLO with \textsc{Pythia} and \textsc{Prospino2} but, unfortunately, it is still too low to explain the observations.

Trying to obtain a model able to account for the excess, we have considered a second point in our GGM scenario with a lighter gluino. The GGM2 point is obtained with $M_s = 400$~TeV, $\tilde B_1^{1/2}=\tilde A_1=309$~TeV, $\tilde B_2^{1/2}=\tilde A_2 =150$~TeV, $\tilde B_3^{1/2}=110$~TeV, $\tilde A_3=270$~TeV and $\tan\beta=9.8$. With these parameters we obtain the spectrum shown in Table \ref{tab2}.

\begin{table}[ht]
\centering
\caption{\label{tab2} SUSY spectrum for the point GGM2. From Ref.~\cite{Barenboim:2015afa}.}
\begin{tabular*}{0.95\textwidth}{@{\extracolsep{\fill}}l c c c c c c c c}
\hline 
{\rm Particle}& $\tilde g$& $\tilde\chi_1^0$ & $\tilde\chi_2^0$ & $\tilde\chi_3^0$  & $\tilde\chi_4^0$ & $\tilde\chi_1^\pm$ & $\tilde\chi_2^\pm$& $\tilde G$ \\
{\rm Mass (GeV)} & 911.4 & 424.9 & 432.7 & 1111.8 & 1117.1 & 425.8 & 1117.2 & $4.8 \times 10^{-9}$ \\
\hline \hline
{\rm Particle}&  $\tilde q_L$ & $\tilde q_R$ & $\tilde b_1$ & $\tilde b_2$ & $\tilde t_1$ &  $\tilde t_2$ & $\tilde l_L$& $\tilde l_R$ \\
{\rm Mass (GeV)} & 2510 & 2470 & 2400 & 2450 & 2250 & 2400 & 5890 & 5360 \\
\hline\hline
{\rm Particle}&  $h$ & $H$ & $A$ & $H^+$ & & & &  \\
{\rm Mass (GeV)} & 118.1 & 1250 & 1250 & 1253 & & & &  \\
\hline
\end{tabular*}
\end{table}
The neutralino mixing matrix in point GGM2 is similar to the corresponding mixing matrix in GGM1, and the BR($\tilde\chi^0_1 \to \tilde G Z$) = 0.94. However, gluino is now much lighter and the gluino-gluino cross section is now $41.6 \pm 7.5$~fb with \textsc{Prospino2} at NLL+NLO, thus we can expect many more gluino pairs to be produced and a larger contribution in the final selection for this point.

\begin{figure}[ht]
  \centering
  \sidecaption
  \includegraphics[width=0.48\linewidth]{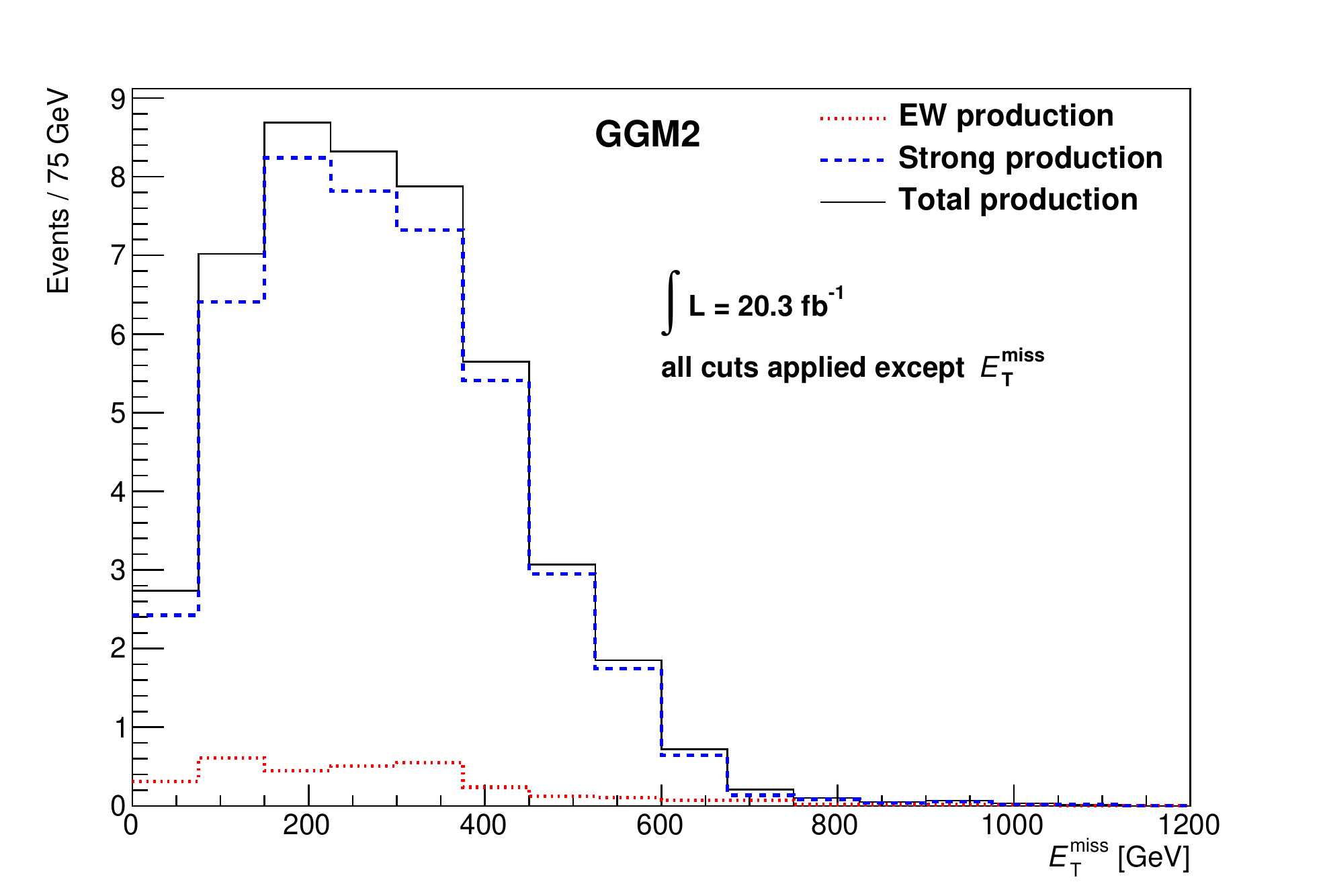}
  \caption{ \met distribution corresponding to the GGM2 point from strong production (solid blue) and electroweak production (dashed red) with the same selection as in Fig.~\ref{GGMHTMET}, after applying the $\HT>600~\GeV$ cut. From Ref.~\cite{Barenboim:2015afa}.}
  \label{GGM2}
  \end{figure}

The simulation for this GGM2 point is presented by the solid blue line in Fig.~\ref{GGMHTMET}. The \HT\ distribution peaks at slightly lower values than in the case of GGM1, due to the slightly lower gluino mass, but it is still enough to overcome the \HT\ cut at 600~\GeV. The \met\ distributions are similar for both GGM points, due to the very similar neutralino masses in both cases. However, in the case of GGM2 the strong production cross section is larger and much more important in relation with the electroweak production: for the GGM2 point, the strong production represents $\sim\!70\%$ of the total cross section. As seen in Fig.~\ref{GGM2}, after applying the selection we obtain an expected signal of $28.0 \pm 4.7$ events, compatible with the excess reported by ATLAS.

We have to emphasise here that it is not difficult to obtain the observed excess for light gluino masses, and a gluino mass in between the two presented examples, $m_{\tilde g} \in [900,1100]~\GeV$, could reproduce the observed signal.  However, these points may be in conflict with direct searches of jets plus \met~\cite{Khachatryan:2015pwa,Aad:2015mia,Aad:2014wea}. There is a tension between this excess and the bounds from gluino searches in jets plus \met. This tension will only be solved by the new data at LHC13.  
Therefore, we have proved that it is indeed possible to construct a supersymmetric model that accommodates the observed excess of lepton pairs on the $Z$~peak. The simulations presented here are only a proof of existence and the final model may be very different. Nevertheless, this model will have to share the main features of the examples that we presented here. 

%%%%%%%%%%%%%%%%%%%%%%%%%%%%%%%%%%%%%%%%%%%%%%%%%%%%%%%%%%%%%%%%%%%%%%%%%%%%%%%%%%%%%%%%%%%%%%%%%%%
\section{Prospects for SUSY at LHC13}

As we have shown in the previous sections, the excess observed in ATLAS, if due to SUSY, requires a gluino of a mass $\sim 1~\TeV$ producing nearly one $Z$~boson per gluino in its decay. This scenario would also require relatively heavy squarks of the first generation with $m_{\tilde q} \gtrsim 2.5~\TeV$. If this is indeed the correct explanation to the observed excess, such light gluinos would be abundantly produced at Run~II in LHC together with other SUSY particles.  Therefore, the results obtained at $pp$ collisions at 13~\TeV\ should confirm or reject this supersymmetric explanation of the ATLAS excess.
 
For point GGM1, the gluino pair production cross section at 8~\TeV\ was $\sigma ( p p \to \tilde g \tilde g)_{\rm LHC8}^{\rm NLL+NLO} = 7.6 \pm 1.3$~fb at NLL+NLO.  Similarly the production cross section at 13~\TeV\ would be $\sigma ( p p \to \tilde g \tilde g)_{\rm LHC13}^{\rm NLL+NLO} = 150 \pm 16$~fb, that is, we would expect to produce 20~times more gluinos at LHC13 for the same integrated luminosity.  Repeating the same exercise with point GGM2, we have a gluino pair production cross section at 8~\TeV\ of $\sigma ( p p \to \tilde g \tilde g)_{\rm LHC8}^{\rm NLL+NLO} = 40 \pm 6$~fb while the production cross section at 13~\TeV\ would be $\sigma ( p p \to \tilde g \tilde g)_{\rm LHC13}^{\rm NLL+NLO} = 550 \pm 70$~fb. In this point, the cross section increases by a factor $\sim 15$ in going from 8 to 13~\TeV\ centre-of-mass energy. 
In any case, for both points this would have unambiguous signatures, both on the $Z$~peak with a scaling of the signal found at LHC8 and in direct searches for gluinos using jets plus missing $E_T$ in the extension of the analysis in Refs.~\cite{Khachatryan:2015pwa,Aad:2015mia,Aad:2014wea}.

Before closing, we should comment on the nature of dark matter in our scenario. As we have seen the signal seems to prefer a non-stable neutralino decaying to a very light gravitino and a $Z$~boson. Under these conditions the neutralino mass has no relation with the dark matter abundance of the Universe and its role as dark matter component would be played by the gravitino. The gravitino mass is not bounded by the observed signal but regardless of its exact mass, unfortunately no signal of dark matter is to be expected in direct search experiments.

%%%%%%%%%%%%%%%%%%%%%%%%%%%%%%%%%%%%%%%%%%%%%%%%%%%%%%%%%%%%%%%%%%%%%%%%%%%%%%%%%%%%%%%%%%%%%%%%%%%
\section{Conclusions}

The ATLAS experiment has announced a $3\sigma$ excess at the $Z$~peak consisting of 29~pairs of leptons observed to be compared with $10.6 \pm 3.2$ expected lepton pairs~\cite{Aad:2015wqa}. No excess outside the $Z$~peak was observed. By trying to explain this signal with SUSY we found that only relatively light gluinos, $m_{\tilde g} \lesssim 1.2~\TeV$, together with a heavy neutralino NLSP of $m_{\tilde \chi_1^0} \gtrsim 400~\GeV$ decaying predominantly to $Z$~boson plus a light gravitino, such that nearly every gluino produces at least one $Z$~boson in its decay chain, could do it. 

The latter is not possible to be reproduced within minimal SUSY models, as mSUGRA, minimal gauge mediation or anomaly mediation. The requirement of a neutralino NLSP decaying to $Z$ plus gravitino points to models of General Gauge mediation as the simplest possibility. It has been shown that a model of this class is able to reproduce the observed signal overcoming all the experimental cuts~\cite{Barenboim:2015afa}. Needless to say, more sophisticated models could also reproduce the signal, however, they will always share the above mentioned features, i.e.\ light gluinos (or heavy particles with a strong production cross section) with an effective ${\cal N}(\tilde g \to Z) \simeq 1$.

%%%%%%%%%%%%%%%%%%%%%%%%%%%%%%%%%%%%%%%%%%%%%
\section*{Acknowledgments}
The author acknowledges support by the Spanish Ministry of Economy and Competitiveness (MINECO) under the project FPA2012-39055-C02-01, by the Generalitat Valenciana through the project PROMETEO~II/2013-017, by the Centro de Excelencia Severo Ochoa SEV-2014-0398 and by the Spanish National Research Council (CSIC) under the JAE-Doc program co-funded by the European Social Fund (ESF). 

%%%%%%%%%%%%%%%%%%%%%%%%%%%%%%%%%%%%%%%%%%%%%
%%%%%%%%%%%%%%%%%%%%%%%%%%%%%%%%%%%%%%%%%%%%%


\begin{thebibliography}{99}
%

\bibitem{susy-rev} H.~P.~Nilles,
  %``Supersymmetry, Supergravity and Particle Physics,''
  {\it Phys.\ Rept.} {\bf 110} (1984) 1; \\
  %%CITATION = PRPLC,110,1;%%
S.~P.~Martin,
  {\it A Supersymmetry primer},  
  in {\it Perspectives on supersymmetry~II} ed.\ G.~L.~Kane, World Scientific (1998) [hep-ph/9709356].
  %%CITATION = HEP-PH/9709356;%%  
  
\bibitem{dm-review} For a review on dark-matter searches in colliders, see e.g.: V.~A.~Mitsou, 
  %``Shedding Light on Dark Matter at Colliders,''
  {\it Int.\ J.\ Mod.\ Phys.} A {\bf 28} (2013) 1330052 [arXiv:1310.1072 [hep-ex]].
  %%CITATION = ARXIV:1310.1072;%% 

\bibitem{Aad:2008zzm}  G.~Aad {\it et al.}  [ATLAS Collaboration],
  %``The ATLAS Experiment at the CERN Large Hadron Collider,''
  JINST {\bf 3} (2008) S08003.
  %%CITATION = JINST,3,S08003;%%

\bibitem{Evans:2008zzb}  L.~Evans and P.~Bryant,
  %``LHC Machine,''
  JINST {\bf 3} (2008) S08001.
  %%CITATION = JINST,3,S08001;%%
  
\bibitem{Aad:2015wqa}  G.~Aad {\it et al.} [ATLAS Collaboration],
  %``Search for supersymmetry in events containing a same-flavour opposite-sign dilepton pair, jets, and large missing transverse momentum in $\sqrt{s}=8$  TeV pp collisions with the ATLAS detector,''
  Eur.\ Phys.\ J.\ C {\bf 75} (2015)  318
   [Erratum-ibid.\ {\bf 75} (2015) 463]   [arXiv:1503.03290 [hep-ex]].
  %%CITATION = ARXIV:1503.03290;%%
  
\bibitem{Barenboim:2015afa}
  G.~Barenboim, J.~Bernabeu, V.~A.~Mitsou, E.~Romero, and O.~Vives,
  ``METing SUSY on the Z peak,''
  arXiv:1503.04184 [hep-ph] (2015).
  %%CITATION = ARXIV:1503.04184;%%

\bibitem{Khachatryan:2015pwa}  V.~Khachatryan {\it et al.} [CMS Collaboration],
  %``Search for Supersymmetry Using Razor Variables in Events with $b$-Tagged Jets in $pp$ Collisions at $\sqrt{s} =$ 8 TeV,''
  Phys.\ Rev.\ D {\bf 91} (2015) 052018  [arXiv:1502.00300 [hep-ex]].
  %%CITATION = doi:10.1103/PhysRevD.91.052018;%%

\bibitem{Aad:2015mia}  G.~Aad {\it et al.} [ATLAS Collaboration],
  %``Search for squarks and gluinos in events with isolated leptons, jets and missing transverse momentum at $\sqrt{s}=8$ TeV with the ATLAS detector,''
  JHEP {\bf 1504} (2015) 116  [arXiv:1501.03555 [hep-ex]].
  %%CITATION = doi:10.1007/JHEP04(2015)116;%%

\bibitem{Aad:2014wea}  G.~Aad {\it et al.}  [ATLAS Collaboration],
  %``Search for squarks and gluinos with the ATLAS detector in final states with jets and missing transverse momentum using $\sqrt{s}=8$ TeV proton--proton collision data,''
  JHEP {\bf 1409} (2014) 176  [arXiv:1405.7875 [hep-ex]].
  %%CITATION = ARXIV:1405.7875;%%

\bibitem{Khachatryan:2014doa}  V.~Khachatryan {\it et al.}  [CMS Collaboration],
  %``Search for top-squark pairs decaying into Higgs or Z bosons in pp collisions at $\sqrt{s}$=8 TeV,''
  Phys.\ Lett.\ B {\bf 736} (2014) 371  [arXiv:1405.3886 [hep-ex]].
  %%CITATION = ARXIV:1405.3886;%%

\bibitem{Aad:2014kra}  G.~Aad {\it et al.}  [ATLAS Collaboration],
  %``Search for top squark pair production in final states with one isolated lepton, jets, and missing transverse momentum in $\sqrt s =$8 TeV $pp$ collisions with the ATLAS detector,''
  JHEP {\bf 1411} (2014) 118  [arXiv:1407.0583 [hep-ex]].
  %%CITATION = ARXIV:1407.0583;%%

\bibitem{Khachatryan:2014mma}  V.~Khachatryan {\it et al.}  [CMS Collaboration],
  %``Searches for electroweak neutralino and chargino production in channels with Higgs, Z, and W bosons in pp collisions at 8 TeV,''
  Phys.\ Rev.\ D {\bf 90} (2014) 092007  [arXiv:1409.3168 [hep-ex]].
  %%CITATION = ARXIV:1409.3168;%%

\bibitem{Aad:2014vma}  G.~Aad {\it et al.}  [ATLAS Collaboration],
  %``Search for direct production of charginos, neutralinos and sleptons in final states with two leptons and missing transverse momentum in $pp$ collisions at $\sqrt{s} =$ 8 TeV with the ATLAS detector,''
  JHEP {\bf 1405} (2014) 071  [arXiv:1403.5294 [hep-ex]].
  %%CITATION = ARXIV:1403.5294;%%

\bibitem{Dine:1993yw}  M.~Dine and A.~E.~Nelson,
  %``Dynamical supersymmetry breaking at low-energies,''
  Phys.\ Rev.\ D {\bf 48} (1993) 1277  [hep-ph/9303230].
  %%CITATION = HEP-PH/9303230;%%

\bibitem{Dine:1994vc}  M.~Dine, A.~E.~Nelson and Y.~Shirman,
  %``Low-energy dynamical supersymmetry breaking simplified,''
  Phys.\ Rev.\ D {\bf 51} (1995) 1362  [hep-ph/9408384].
  %%CITATION = HEP-PH/9408384;%%

\bibitem{Dine:1995ag}  M.~Dine, A.~E.~Nelson, Y.~Nir and Y.~Shirman,
  %``New tools for low-energy dynamical supersymmetry breaking,''
  Phys.\ Rev.\ D {\bf 53} (1996) 2658  [hep-ph/9507378].
  %%CITATION = HEP-PH/9507378;%%

\bibitem{Giudice:1998bp}  G.~F.~Giudice and R.~Rattazzi,
  %``Theories with gauge mediated supersymmetry breaking,''
  Phys.\ Rept.\  {\bf 322} (1999) 419  [hep-ph/9801271].
  %%CITATION = HEP-PH/9801271;%%

\bibitem{Feng:2004mt}  J.~L.~Feng, S.~Su and F.~Takayama,
  %``Supergravity with a gravitino LSP,''
  Phys.\ Rev.\ D {\bf 70} (2004) 075019  [hep-ph/0404231].
  %%CITATION = HEP-PH/0404231;%%

\bibitem{Moroi:1995fs}  T.~Moroi,
 ``Effects of the gravitino on the inflationary universe,''
  hep-ph/9503210 (1995).
  %%CITATION = HEP-PH/9503210;%%

\bibitem{Ellis:2003dn}  J.~R.~Ellis, K.~A.~Olive, Y.~Santoso and V.~C.~Spanos,
  %``Gravitino dark matter in the CMSSM,''
  Phys.\ Lett.\ B {\bf 588} (2004) 7  [hep-ph/0312262].
  %%CITATION = HEP-PH/0312262;%%

\bibitem{Meade:2008wd}  P.~Meade, N.~Seiberg and D.~Shih,
  %``General Gauge Mediation,''
  Prog.\ Theor.\ Phys.\ Suppl.\  {\bf 177} (2009) 143  [arXiv:0801.3278 [hep-ph]].
  %%CITATION = ARXIV:0801.3278;%%

\bibitem{Buican:2008ws}  M.~Buican, P.~Meade, N.~Seiberg and D.~Shih,
  %``Exploring General Gauge Mediation,''
  JHEP {\bf 0903} (2009) 016  [arXiv:0812.3668 [hep-ph]].
  %%CITATION = ARXIV:0812.3668;%%

\bibitem{Carpenter:2008he}  L.~M.~Carpenter,
  ``Surveying the Phenomenology of General Gauge Mediation,''
  arXiv:0812.2051 [hep-ph] (2008).
  %%CITATION = ARXIV:0812.2051;%%

\bibitem{Rajaraman:2009ga}  A.~Rajaraman, Y.~Shirman, J.~Smidt and F.~Yu,
  %``Parameter Space of General Gauge Mediation,''
  Phys.\ Lett.\ B {\bf 678} (2009) 367  [arXiv:0903.0668 [hep-ph]].
  %%CITATION = ARXIV:0903.0668;%%

\bibitem{Thalapillil:2010ek}  A.~M.~Thalapillil,
  %``Low-energy Observables and General Gauge Mediation in the MSSM and NMSSM,''
  JHEP {\bf 1106} (2011) 059  [arXiv:1012.4829 [hep-ph]].
  %%CITATION = ARXIV:1012.4829;%%

\bibitem{Kats:2011qh}  Y.~Kats, P.~Meade, M.~Reece and D.~Shih,
  %``The Status of GMSB After 1/fb at the LHC,''
  JHEP {\bf 1202} (2012) 115  [arXiv:1110.6444 [hep-ph]].
  %%CITATION = ARXIV:1110.6444;%%

\bibitem{Ruderman:2011vv}  J.~T.~Ruderman and D.~Shih,
  %``General Neutralino NLSPs at the Early LHC,''
  JHEP {\bf 1208} (2012) 159  [arXiv:1103.6083 [hep-ph]].
  %%CITATION = ARXIV:1103.6083;%%

\bibitem{Porod:2003um}  W.~Porod,
  %``SPheno, a program for calculating supersymmetric spectra, SUSY particle decays and SUSY particle production at e+ e- colliders,''
  Comput.\ Phys.\ Commun.\  {\bf 153} (2003) 275  [hep-ph/0301101].
  %%CITATION = HEP-PH/0301101;%%

\bibitem{Porod:2011nf}  W.~Porod and F.~Staub,
  %``SPheno 3.1: Extensions including flavour, CP-phases and models beyond the MSSM,''
  Comput.\ Phys.\ Commun.\  {\bf 183} (2012) 2458  [arXiv:1104.1573 [hep-ph]].
  %%CITATION = ARXIV:1104.1573;%%

\bibitem{Dine:2007xi}  M.~Dine, N.~Seiberg and S.~Thomas,
  %``Higgs physics as a window beyond the MSSM (BMSSM),''
  Phys.\ Rev.\ D {\bf 76} (2007) 095004  [arXiv:0707.0005 [hep-ph]].
  %%CITATION = ARXIV:0707.0005;%%

\bibitem{0710.3820}  T.~Sjostrand, S.~Mrenna and P.~Z.~Skands,
  %``A Brief Introduction to PYTHIA 8.1,''
  Comput.\ Phys.\ Commun.\  {\bf 178} (2008) 852 [arXiv:0710.3820 [hep-ph]].
  %%CITATION = ARXIV:0710.3820;%%  

\bibitem{Beenakker:1996ch}  W.~Beenakker, R.~Hopker, M.~Spira and P.~M.~Zerwas,
  %``Squark and gluino production at hadron colliders,''
  Nucl.\ Phys.\ B {\bf 492} (1997) 51  [hep-ph/9610490].
  %%CITATION = HEP-PH/9610490;%%

\bibitem{Beenakker:1997ut}  W.~Beenakker, M.~Kramer, T.~Plehn, M.~Spira and P.~M.~Zerwas,
  %``Stop production at hadron colliders,''
  Nucl.\ Phys.\ B {\bf 515} (1998) 3  [hep-ph/9710451].
  %%CITATION = HEP-PH/9710451;%%

\bibitem{Beenakker:1999xh}  W.~Beenakker, M.~Klasen, M.~Kramer, T.~Plehn, M.~Spira and P.~M.~Zerwas,
  %``The Production of charginos / neutralinos and sleptons at hadron colliders,''
  Phys.\ Rev.\ Lett.\  {\bf 83} (1999) 3780   [Erratum-ibid.\  {\bf 100} (2008) 029901]  [hep-ph/9906298].
  %%CITATION = HEP-PH/9906298;%%

\bibitem{deFavereau:2013fsa}  J.~de Favereau {\it et al.}  [DELPHES 3 Collaboration],
  %``DELPHES 3, A modular framework for fast simulation of a generic collider experiment,''
  JHEP {\bf 1402} (2014) 057  [arXiv:1307.6346 [hep-ex]].
  %%CITATION = ARXIV:1307.6346;%%



\end{thebibliography}
\end{document}